# PARISROC, a Photomultiplier Array Integrated Read Out Chip.

S. Conforti Di Lorenzo*, J.E.Campagne, F. Dulucq*,C. de La Taille*, G. Martin-Chassard*, M. El Berni.
LAL/IN2P3, Laboratoire de l'Accélérateur Linéaire, Université Paris-Sud 11, UMR 8607, 91898 Orsay cedex

W. Wei. IHEP, Beijing, China.

*Abstract*–PARISROC is a complete read out chip, in AMS SiGe 0.35 μm technology, for photomultipliers array. It allows triggerless acquisition for next generation neutrino experiments and it belongs to an R&D program funded by the French national agency for research (ANR) called PMm$^2$: "Innovative electronics for photodetectors array used in High Energy Physics and Astroparticles" (ref.ANR-06-BLAN-0186).

The ASIC (Application Specific Integrated Circuit) integrates 16 independent and auto triggered channels with variable gain and provides charge and time measurement by a Wilkinson ADC (Analog to Digital Converter) and a 24-bit Counter. The charge measurement should be performed from 1 up to 300 photo-electrons (p.e.) with a good linearity. The time measurement allowed to a coarse time with a 24-bit counter at 10 MHz and a fine time on a 100ns ramp to achieve a resolution of 1 ns. The ASIC sends out only the relevant data through network cables to the central data storage. This paper describes the front-end electronics ASIC called PARISROC.

## I. INTRODUCTION

The next generation of proton decay and neutrino experiments, the post-SuperKamiokande detectors as those that will take place in megaton size water tanks, will require very large surfaces of photodetection, even with large hemispherical photomultiplier tubes (PMT), and a large volume of data: the expected number of channels should reach hundreds of thousands. The PMm$^2$ project [1] proposes to segment the large surface of photodetection in macro pixel consisting of an array (2 m × 2m) of 16 photomultipliers (16 PMTs as SuperKamiokande module structure) connected to an autonomous front-end electronics (Fig.1) and powered by a common High Voltage (HV). This R&D involves three French laboratories (LAL Orsay, LAPP Annecy, IPN Orsay) and ULB Bruxells for the DAQ. It is funded by the French National Agency for Research (ANR) under the reference ANR-06-BLAN-0186.

The micro-electronics group's (OMEGA from the LAL at Orsay) is in charge of the design and tests of the readout chip named PARISROC which stands for Photomultiplier ARrray Integrated in Si-Ge Read Out Chip.

The detectors are large tanks covered by a significant number of large photomultipliers (20-inch); the next generation neutrino experiments will require a bigger surface of photo detection and thus more photomultipliers. As a consequence the total cost has an important relief [2].

The project proposes to reduce the costs using:

1. 12-inch PMts, instead of 20-inch, with an improved cost per unit of surface area and detected p.e (due to a different industrial fabrication of the PMTs);
2. A smaller number of electronics, thanks to the 16 PMTs macropixel with a common electronics;
3. The designing of the ASIC as a System on a Chip (SoC) that processing the analog signal up to digitization (cost improvement only in a large scale of production);
4. A common High Voltage for the 16 PMTs so a reduced number of underwater cables, cables that are also used to brought the DATA to the surface;
5. The front-end closed to the PMTs that allow a suppression of underwater connectors.

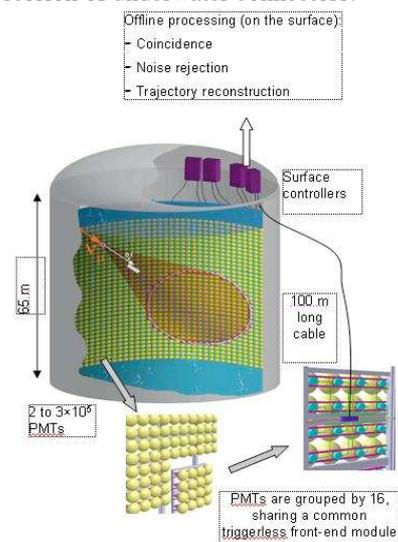

Fig .1. Principal of PMm$^2$ proposal for megaton scale Cerenkov water tank with 16 PMTs array shown in the bottom side of the picture and are emphasized the surface controller and the 100m of cable.

The general principle of PMm$^2$ project is that the ASIC and a FPGA (Field-Programmable gate array) manage the dialog between the PMTs and the surface controller (Fig. 2).

Through the 100 m of data cable (used also to transmit the slow control, the power supply, the 10 MHz reference clock) the digitized data are transmitted with a dedicated protocol at a rate of 5 Mbit/s. The data are processed at the surface level on a network: the interface with the front end is carried out by a surface controller [2].

Alternative options may be chosen considering an analysis of the risks of this full underwater strategy. One option is that the Front-End electronics can be used in a traditional schema with the electronic "in surface". PARISROC can be integrated perfectly in a surface scheme.

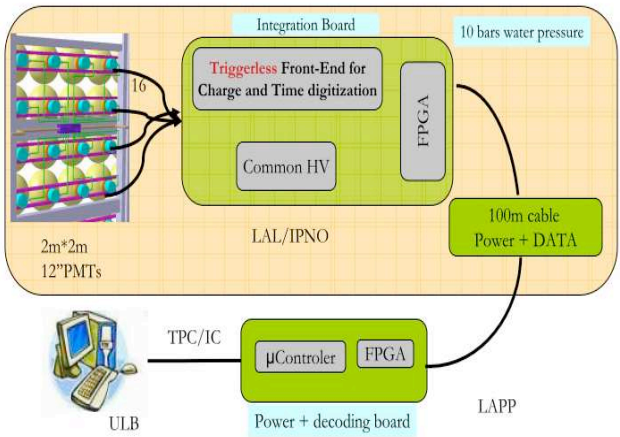

Fig. 2. Principle of the PMm$^2$ project. The main blocks are shown: the integration board that receive the PMT signal and transmit by 100 m of cable the 52 bits of DATA (for each channel) to the surface controller.

## II. REQUIREMENTS

The physics events researched in the detectors, as SuperKamiokande one, produce Cerenkov light that is spread over the PMTs. This kind of experiences we are looking for rare events and the number of pe per deposit MeV in the water is of the order of 10pe/MeV and may be viewed by O(10000) PMTs. So, for few MeV events, the small number of p.e is spread over a large number of PMTs and as consequence it is necessary being fully efficient to detect a single photo-electron (p.e). For large energy events (atmospheric or beam neutrinos) it is shown by simulation that the dynamic range for a single PMT should cover up to few hundred p.e (~ 300pe). This induces an electronic that covers a dynamic range from 1 to 300 pe and a trigger at 1/3 p.e.

All the type of events considered must be registered without any direct external trigger, this is called 'triggerless mode' and requires an electronic with autotrigger channels.

A precise time stamp of each event is required to reconstruct the topology of the events and so to synchronize the events among PMTs in each array and among the different arrays. This aspect brought to a requirement: an electronic with full independent channels.

The most demanding in term of timing is the vertex reconstruction that needs typically 1 ns resolution considering the single electron jitter around 3 ns (for 12-inch PMT).

The mean PMT gain is about $3 \times 10^6$ but a spread is foreseen among the PMTs of the same array as a single HV is used. It was estimate from the production of the Auger experiment PMTs that the gain dispersion, at a given voltage is such that, the ratio between the highest and the lowest gain, is not more than 12 [2]. It is possible for the manufacturer to sort the PMTs, at a reasonable cost, when they are produced at a very large scale, with a gain ratio reduced to 6 in a batch of 16 PMTs. To compensate this not homogeneity a preamplifier with a variable and adjustable gain is required.

To summarize, the electronic requirements are: 1pe full efficiency, triggerless acquisition, 1ns of time resolution, 300 pe of dynamic range, high granularity, scalability, low cost, independent channels, water-tight, common high voltage, only digitized data to be send to the surface (DATA + VCC).

## III. PARISROC ARCHITECTURE

### A. Global architecture.

The ASIC PARISROC (Fig. 3) is composed of 16 analog channels managed by a common digital part.

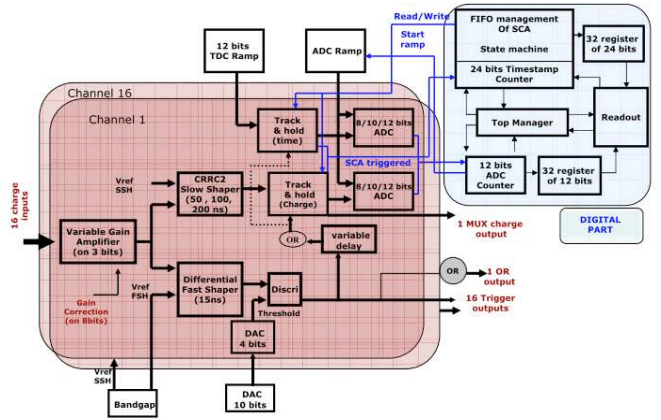

Fig. 3. PARISROC global schematic.

Each analog channel (Fig. 4) is made of a voltage preamplifier with variable and adjustable gain. The variable gain is common for all channels and it can change, thanks to the input variable capacitance, on 3 bits. The gain is also tuneable channel by channel to adjust the input PMTs gain non-homogeneity, thanks to the switched feedback capacitance, on 8 bits.

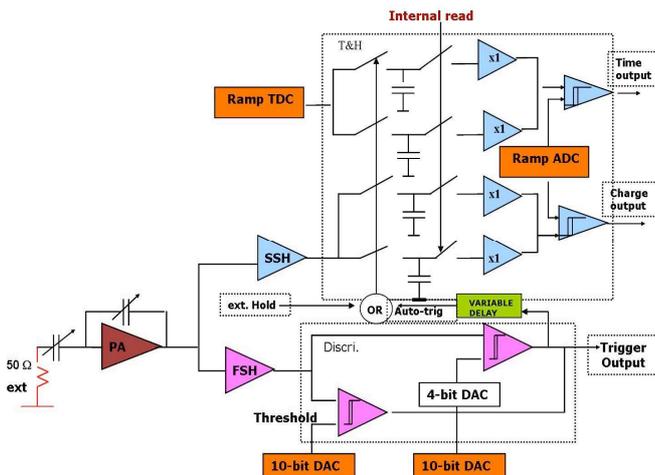

Fig. 4. PARISROC one channel analog part schematic. In the picture are emphasized, on the left the preamplifier with variable input and feedback capacitances; on the top, the Track &Hold block with the two capacitances to memorize charge and time measurements and the two ADC, with a common ADC ramp, to convert the analog values.

The preamplifier is followed by a slow channel, for the charge measurement, in parallel with a fast channel, for the trigger output. The slow channel is made by a slow shaper followed by an analogue memory, with a depth of 2, to provide a linear charge measurement up to 50 pC; this charge is converted by a Wilkinson ADC (8, 10, 12-bit). One follower OTA is added to deliver an analogue multiplexed charge measurement. The fast channel consists in a fast shaper (15 ns) followed by 2 low offset discriminators, to auto-trig down to 50 fC. The thresholds are loaded by 2 internal 10-bit DACs (Digital to Analog Converter) common for the 16 channels and an individual 4bit DAC for one discriminator. The 2 discriminator outputs are multiplexed to provide only 16 trigger outputs. Each output trigger is latched to hold the state of the response until the end of the clock cycle. It is also delayed to open the hold switch at the maximum of the slow shaper. An "OR" of the 16 trigger gives a 17th output. For each channel, a fine time measurement is made by an analogue memory, with depth of 2, which samples a 12-bit ramp, common for all channels, at the same time of the charge. This time is then converted by a Wilkinson ADC. The two ADC discriminators have a common ramp, of 8, 10, 12 bits, as threshold to convert the charge and the fine time. In addition a bandgap bloc provides all voltage references.

### B. Digital part.

The digital part of PARISROC is built around 4 modules which are acquisition, conversion, readout and top manager [3]. Actually, PARISROC is based on 2 memories. During acquisition, discriminated analog signals are stored into an analog memory (the SCA: switched capacitor array). The analog to digital conversion module converts analog charges and times from SCA into 12 bits digital values. These digital values are saved into registers (RAM). At the end of the cycle, the RAM is readout by an external system. The block diagram is given below on Fig. 5.

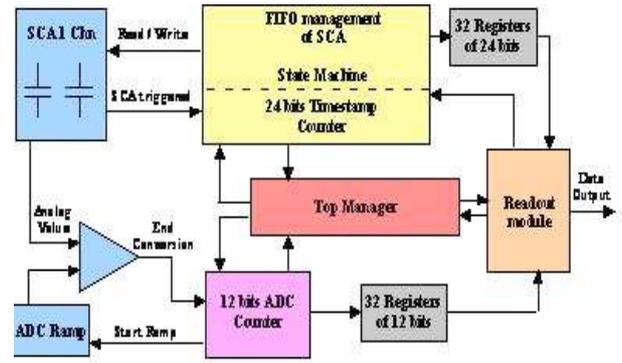

Fig. 5. Block diagram of the digital part.

The sequence is made thanks to the top manager module which controls the 3 other ones. When 1 or more channels are hit, it starts ADC conversion and then the readout of digitized data. The maximum cycle length is about 200 $\mu$s. During conversion and readout, acquisition is never stopped. It means that discriminated analog signals can be stored in the SCA at any time.

The first module in the sequence is the acquisition which is dedicated to charge and fine time measurements. It manages the SCA where charge and fine time are stored as a voltage like. It also integrates the coarse time measurement thanks to a 24-bit gray counter with a resolution of 100 ns. Each channel has a depth of 2 for the SCA and they are managed individually. Besides, SCA is treated like a FIFO memory: analog voltage can be written, read and erased from this memory.

Then, the conversion module converts analog values stored in the SCA (charge and fine time) in digital ones thanks to a 12-bit Wilkinson ADC. The counter clock frequency is 40 MHz, it implies a maximum ADC conversion time of 103 $\mu$s when it overflows. This module makes 32 conversions in 1 run (16 charges and 16 fine times).

Finally, the readout module permits to empty all the registers to an external system. As it will only transfer hit channels, this module tag each frame with its channel number: it works as a selective readout. The pattern used is composed of 4 data: 4-bit channel number, 24-bit coarse time, 12-bit charge and 12-bit fine time. The total length of one frame is 52 bits. The maximum readout time appears when all channels are hit. About 832 bits of data are transferred to the concentrator with a 10 MHz clock: the readout takes about 100 $\mu$s with 1 $\mu$s between 2 frames.

### C. PARISROC layout

Figure 6 presents the PARISROC layout [4]. The circuit has been designed in AMS SiGe 0.35 $\mu$m technology [5]. The die has a surface of 17 mm$^2$ (5 mm X 3.4 mm) and will be package in CQFP160 case.

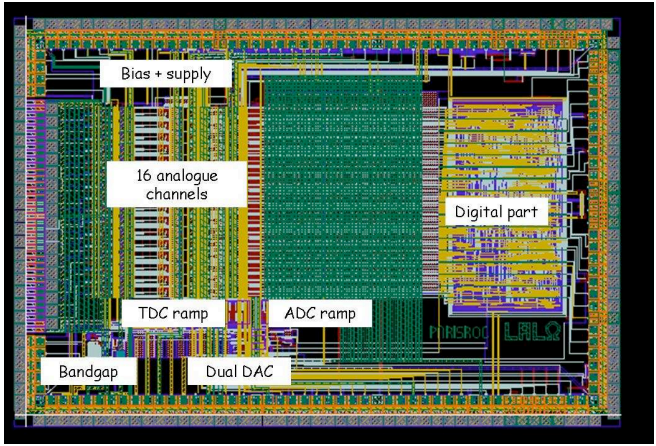

Fig. 6. PARISROC layout.

The PARISROC has been submitted in June 2008; a first batch of 6 ASICs has been produced and received in December 2008 (a second batch of 14 ASICs in May 2009).

## IV. MEASUREMENTS AND SIMULATIONS.

### A. General tests.

A dedicated test board has been designed and realized for testing the ASIC (Fig. 7). Its aim is to allow the characterization of the chip and the communication between photomultipliers and ASIC. This is possible thanks to a dedicated Labview program that allows to send the ASIC configuration (slow control parameters, ASIC parameters, etc) and to receive the output bits via an USB cable connected to the test board. The Labview is developed by the LAL "Tests group".

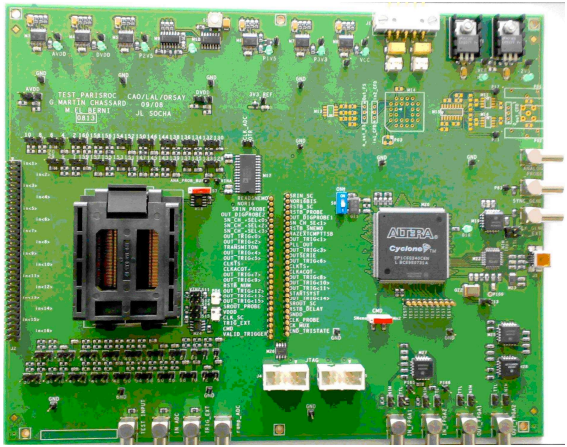

Fig. 7. Test Board.

A signal generator is used to create the input charge injected in the ASIC. The signal injected is similar, as possible, to the PMT signal. In Fig. 8 is represented the input signal and its characteristics. The input signal, used in measurements and simulation, is a triangle signal with 5 ns rise and fall time and 5 ns of duration. This current signal is sent to an external resistor (50 Ohms) and varies from 0 to 5 mA in order to simulate a PMT charge from 0 to 50 pC which represents 0 to 300 p.e. when the PM gain is $10^6$.

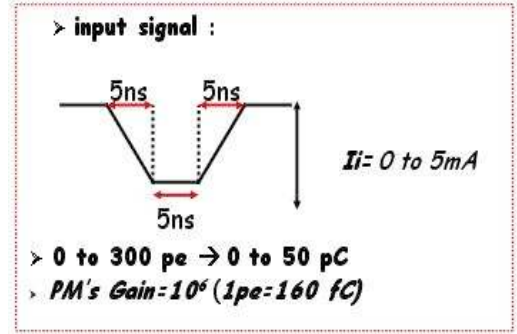

Fig. 8. Input signal used for measurements and simulations.

Table I lists the simulation and measurements results for the three main blocks of the analog part: Preamplifier, Slow shaper and Fast shaper.

TABLE I. ANALOG PART MEASUREMENTS

|  | Preamplifier Gain P.A.=8 Meas./Sim. | Slow Shaper RC=50ns Meas./Sim. | Fast Shaper Meas./Sim. |
|---|---|---|---|
| Voltage (1 p.e.) | 5mV/5.43mV | 12mV/19mV | 30mV/39mV |
| Rms noise | 1mV/468µV | 4mV/2.3mV | 2.5mV/2.4mV |
| Noise in p.e. | 0.2/0.086 | 0.3/0.125 | 0.08/0.06 |
| SNR | 5/12 | 3/8 | 12/16 |

There is a good agreement between measurements and simulations in analog part results except for the noise values. To characterize the noise, the Signal to Noise Ratio (SNR) is calculated with reference to the MIP (1 p.e.). The noise differences are immediately evident: an additional low frequency noise is present in measurement (is now under investigation even if it is supposed to be tied to the power supply noise). A small difference has been noticed in measurement without the USB cable that allowed the communication between the test board and the Labwiev program: an rms noise value of 660 µV (0.132 p.e.) for preamplifier and so a SNR value of 8 (slightly nearer to the simulation value of 12).

Another important characteristic is the linearity. The preamplifier linearity in function of variable feedback capacitor value with an input charge of 10 p.e. and with residuals from -2.5 to 1.35 % is represented on Fig. 9. The gain adjustment linearity is good at 2% on 8 bits.

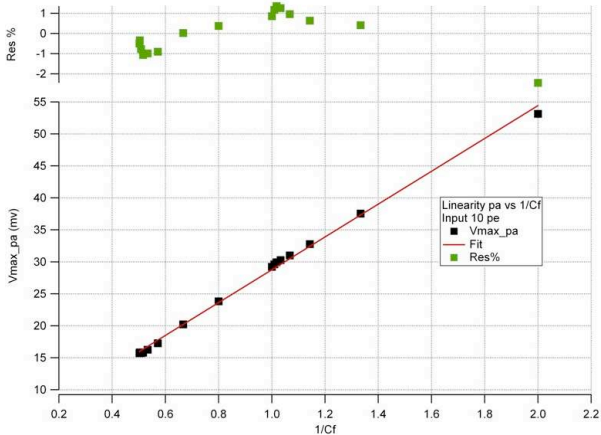

Fig. 9. Preamplifier linearity vs feedback capacitor value.

Figure 10 represents the slow shaper linearity for a time constant of 50 ns and a preamplifier gain of 8. The slow shaper output voltage in function of the input injected charge is plotted. Good linearity performances are obtained with residuals better than ± 1%.

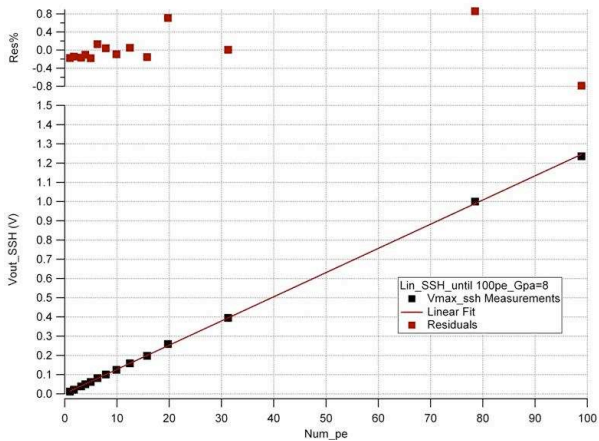

Fig. 10. Slow shaper linearity; τ=50 ns and Gpa=8.

In order to investigate the homogeneity among the whole chip, essential for a multichannel ASIC, for different preamplifier gains are plotted the maximum voltage values for all channels. On Fig. 11 is given the gain uniformity. A good dispersion of 0.5%, 1.4% and 1.2% have respectively been obtained for gain 8, 4 and 2. This represents a goal for the ASIC.

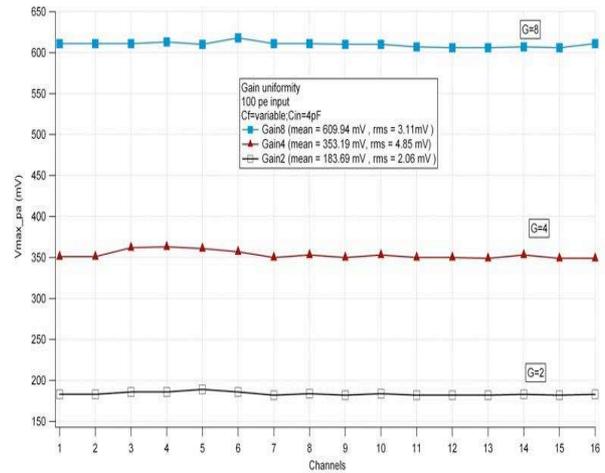

Fig. 11. Gain uniformity for Gpa = 8, 4, 2.

### B. DAC Linearity

The DAC linearity (Fig. 12) has been measured and it consists in measuring the voltage DAC (Vdac) amplitude obtained for different DAC register values. Figure 12 gives the evolution of Vdac as a function of the register for the DAC and the residuals with values from -0.1% to 0.1%.

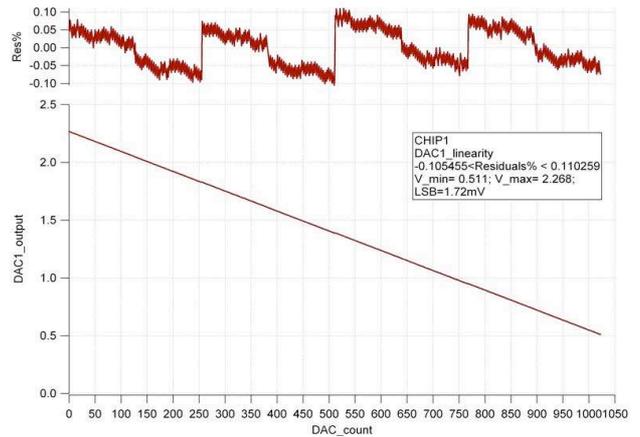

Fig. 12. DAC Linearity.

### C. Trigger Output.

The trigger output behaviour was studied scanning the threshold for different injected charges. At first no charge was injected which corresponds to measure the fast shaper pedestal. The result is represented on Fig. 13 for each channel. The 16 curves (called s-curves because of their shape) are superimposed that meaning good homogeneity. The spread is of one DAC count (LSB DAC=1.78 mV) equivalent to 0.06 p.e.

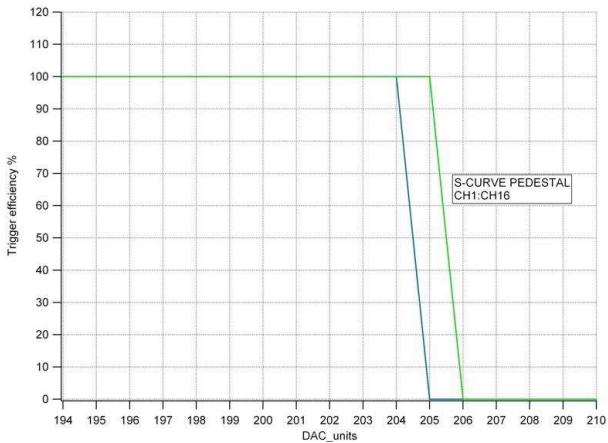

Fig. 13. Pedestal S-Curves for channel 1 to 16.

The trigger efficiency was then measured for a fixed injected charge of 10 p.e. On Fig. 14 are represented the S-curves obtained with 200 measurements of the trigger for all channels varying the threshold. The homogeneity is proved by a spread of 7 DAC units (0.4 p.e).

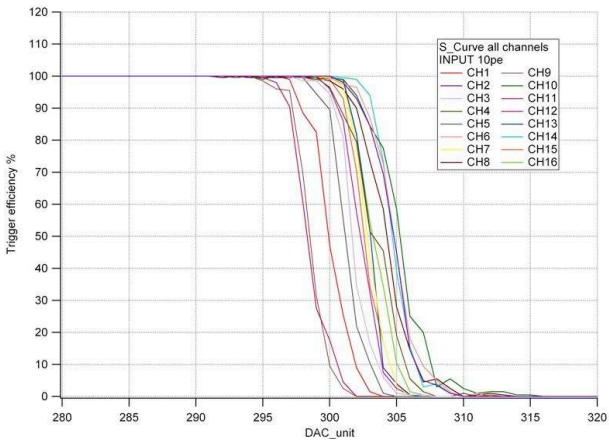

Fig. 14. S-Curve for input of 10 p.e. for channel 1 to 16.

The trigger output is studied also by scanning the threshold for a fixed channel and changing the injected charge. Figure 15 shows the trigger efficiency versus the DAC unit with an injected charge from 0 to 300 p.e. and on Fig. 16 is plotted the threshold versus the injected charge but only until 3 pe.

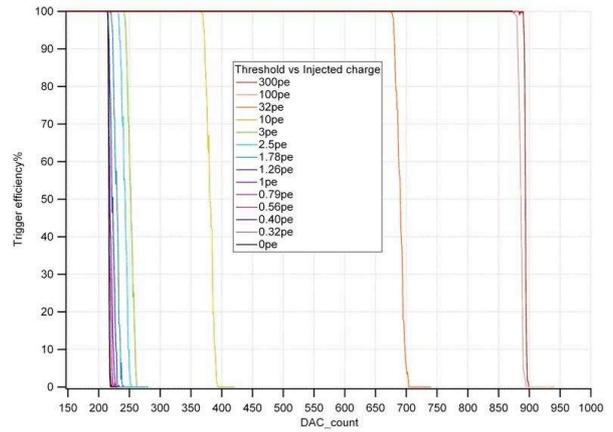

Fig. 15. Trigger efficiency vs DAC count up to 300 p.e.

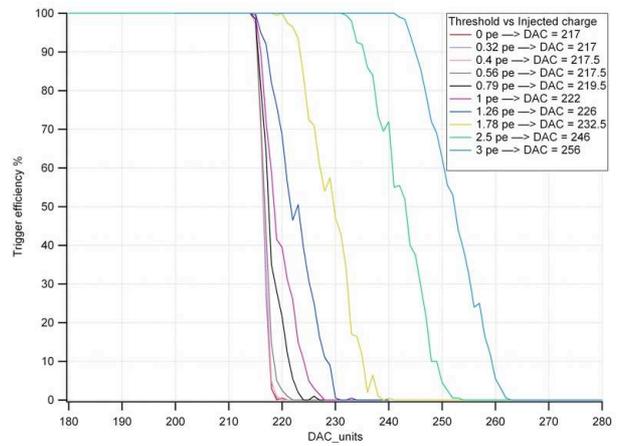

Fig. 16. Trigger efficiency vs DAC count until 3 p.e.

In Fig. 17 are plotted the 50% trigger efficiency values, extracted from the plot in Fig. 16, converted in mV versus the injected charges. A noise of 10 fC has been extrapolated. Therefore the threshold is limited to 10 σ noise due to the discriminator coupling.

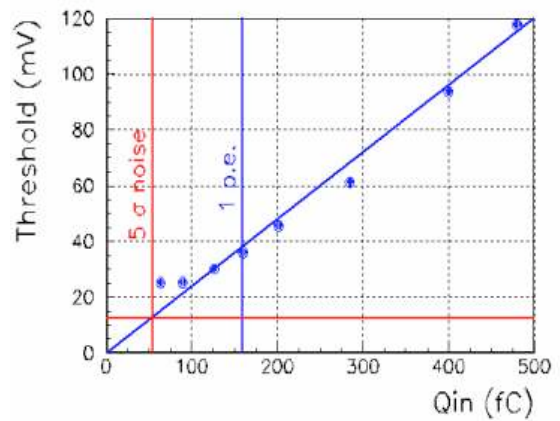

Fig. 17. Threshold vs injected charge until 500 fC.

*D. ADC.*

The ADC performance has been studied alone and with the whole chain. Injecting to the ADC input directly DC voltages by the internal DAC (in order to have a voltage level as stable as possible) the ADC values for all channels have been measured. The measurement is repeated 10000 times for each channel and in the first panel of the Labview window (Fig. 18) the minimal, maximal and mean values, over all acquisitions, for each channel are plotted. In the second panel there is the rms charge value versus channel number with a value in the range 0.5 to 1 ADC unit. Finally the third panel shows an example of charge amplitude distribution for a single channel; a spread of 5 ADC counts is obtained.

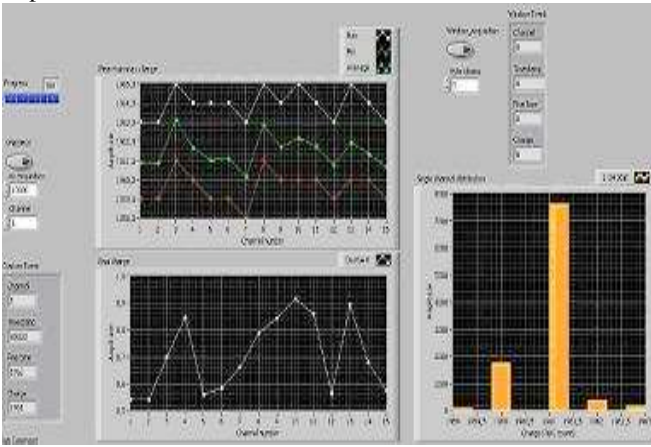

Fig. 18. ADC measurements with DC input 1.45V.

The ADC is suited to a multichannel conversion so the uniformity and linearity are studied in order to characterize the ADC behaviour. On Fig. 19 is represented the ADC transfer function for the 10-bit ADC versus the input voltage level. All channels are represented and have plots superimposed.

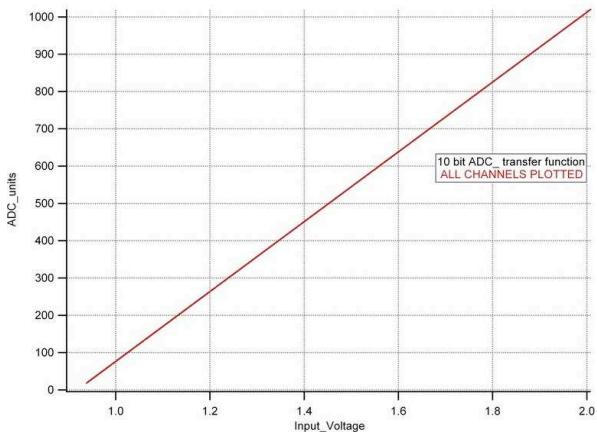

Fig. 19. 10-bit ADC transfer function vs input charge.

This plot shows the good ADC uniformity among the 16 channels. In Fig. 20 is shown the 12-bit ADC linearity plots with the 25 measurements made at each input voltage level. The average ADC count value is plotted versus the input signal. The residuals from -1.5 to 0.9 ADC units for the 12-bit ADC; from -0.5 to 0.4 for the 10-bit ADC and from -0.5 to 0.5 for the 8-bit ADC prove the good ADC behaviour in terms of Integral non linearity.

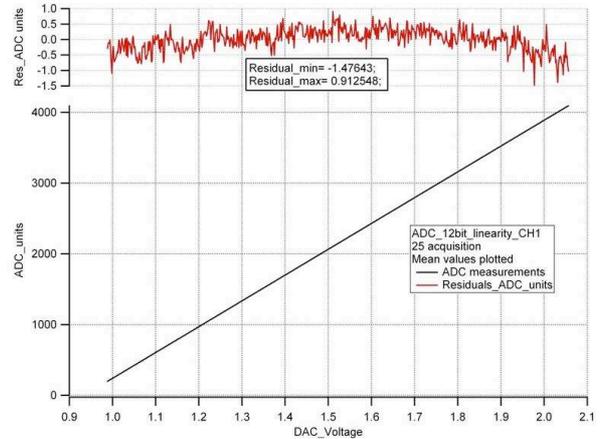

Fig. 20. 12-bit ADC linearity.

Once the ADC performances have been tested separately, the measurements are performed on the complete chain. The results of the input signal auto triggered, held in the T&H and converted in the ADC are illustrated in Fig. 21 where are plotted the 10-bit ADC counts in function of the variable input charge (up to 50 p.e). A nice linearity of 1.4% and a noise of 6 ADC units are obtained.

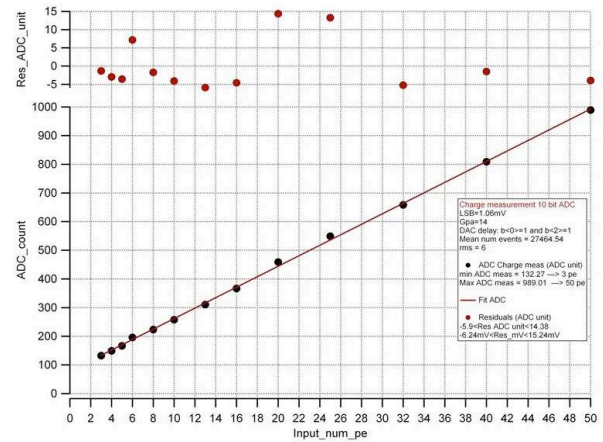

Fig. 21. 10-bit ADC linearity.

## V. CONCLUSION

Good overall performances of the chip PARISROC are obtained: auto trigger signal and digitalization of DATA. Good uniformity and linearity although strange noise performance, due to 10 MHz clock noise and a low frequency noise, now under investigation. A second version of the chip will be submitted in November 2009 with an increasing of the dynamic range thanks to 2 preamplifier gains: high gain and low gain preamplifier; 8/9/10 bits ADC to reduce the p.e. loss below 1% level in case of 5 kHz dark current per PMT and a double fine TAC.